\documentclass[aps,prl,twocolumn]{revtex4}
\usepackage{epsf}
 
\begin{document}

\title{Effect of the on-site Coulomb repulsion on
       superconductivity\\ in the boson-fermion model}
 
\author{Tadeusz\ Doma\'nski}
\affiliation{Institute of Physics, M.\ Curie Sk\l odowska University,
         20-031 Lublin, Poland}

\date{\today}

\begin{abstract}
We study the influence of the repulsive Coulomb interactions
on thermodynamic properties of the boson fermion model
with an anisotropic ($d$-wave, and extended $s$-wave) 
order parameter. Superconductivity is induced in this 
model from the anisotropic charge exchange interaction 
between the conduction band fermions (electrons or holes)
and the immobile hard-core bosons (the localized electron 
pairs). The on-site Coulomb repulsion competes with this 
pairing interaction and hence is expected to have a detrimental
influence on superconductivity. We analyze this effect 
in some detail considering the two opposite limits of: 
the weak and strong repulsion. A possible crossover between 
both these regimes is also discussed.
\end{abstract}

\maketitle
  
\narrowtext
\section{Introduction}
The boson fermion (BF) model describes a mixture of the narrow 
band fermions coupled to a system of the composite hard-core bosons.
Initially, this type of an effective Hamiltonian has been invented 
for a system of itinerant electrons interacting with the local lattice 
deformations in the crossover regime, between adiabatic and 
antiadiabatic limits \cite{Ranninger-85}. Later, the same model 
has been independently considered by a number of authors 
\cite{Eliashberg-87,Friedberg-89,Ioffe-89,Ranninger-95} 
as a possible scenario for a mechanism of the high temperature 
superconductivity (HTSC). There are also attempts to apply 
a similar BF model to explain certain aspects of the Bose 
condensed atoms of the alkali metals \cite{Holland-01}.

This model reveals a rich physics both in its: normal 
phase and the broken symmetry superconducting/superfluid 
state. As shown in the mean field type studies 
\cite{Friedberg-89,Ranninger-95} there is a characteristic 
temperature $T_{c}$ bellow which fermions are driven to 
the superconducting phase and simultaneously bosons start 
to Bose-condense. This result has been confirmed 
(neglecting the hardcore nature of bosons) by means 
of the shielded potential approximation \cite{Kostyrko-96} 
and with a help of the renormalization group approach 
\cite{Domanski-01}. Moreover, when approaching the critical 
temperature from above, the pairing-wise correlations start 
to manifest themselves strongly. In particular, they may give 
rise to a formation of the pseudogap in the fermion spectrum. 
This effect, known experimentally from a variety of measurements 
(see e.g.\ review paper \cite{Timusk-99}), provides a firm argument
for application of this model to describe the HTSC materials.

Pseudogap formation and its variation with a lowering temperature 
has been carefully investigated for the BF model using:
a) the selfconsistent perturbative treatment of the 
   boson-fermion coupling \cite{perturbative,Ren-98}, 
b) the perturbative treatment of the kinetic processes 
   (a l\`a Hubbard I for the fermion hopping) with respect 
   to the exact solution of this model in its atomic limit 
   \cite{Domanski-98},  
c) the dynamical mean field theory (DMFT) equations 
   which have been selfconsistently solved within 
   the noncrossing approximation for the auxiliary 
   impurity problem \cite{DMFT}, 
d) and the renormalization group technique \cite{Domanski-01}.

Many experimental data, especially the angle resolved photoemission 
spectroscopy \cite{ARPES}, seem to suggest the anisotropic $d$-wave 
type structure of both: the pseudogap and true superconducting gap. 
However, there are also known some measurements, for instance the 
$c$-axis Josephson tunneling \cite{Kouznetsov-97} and the photoemission 
spectroscopy on Bi$_{2}$Sr$_{2}$CaCu$_{2}$O$_{8+\delta}$ \cite{Ma-95}, 
which provide the strong arguments for a nonzero $s$-wave ingredient of 
the order parameter. In a most realistic situation one can expect
that the order parameter of the HTSC cuprates acquires a mixed $s+d$ 
or $s+$ i$d$ symmetry. Possibility for an appearance of the mixed 
symmetry superconducting phase has been theoretically explored on 
quite the general grounds using a 2-dimensional electron system 
with the anisotropic potential $V=V_{s}+V_{d}$ of the arbitrary 
(from weak to strong) attraction strength \cite{Musaelian-96}.
So far, most of the studies of superconductivity within the BF
model have been performed for the isotropic pairing interaction.
Some attempts to analyze the $d$-pairing superconductivity together
with a microscopic justification for introducing the BF type
Hamiltonian can be found in the paper by Geshkenbein {\em et al}
\cite{Geshkenbeim-97}. Very recently, a more formal way has been
explored by Micnas {\em et al} \cite{Micnas-01}.

In this paper we shall investigate various kinds of the superconducting
phase induced by an anisotropic potential of the BF model in a presence 
of the Coulomb interactions between fermions. For simplicity we shall 
concentrate only on a case of the on-site repulsion $U>0$, where 
$U=(ii|\frac{e^{2}}{|{\bf r}|}|ii)$ in the Wannier representation. 
In general, one expects that the on-site repulsion (which prevents 
fermions from forming the local pairs) would compete with the correlations 
induced by the boson-fermion coupling (this interaction is a driving force
for the pairing in the BF model and is responsible for inducing 
the pseudogap at temperatures $T^{*} > T > T_{c}$ and the true 
superconducting gap when $T \leq T_{c}$). We shall address the 
following question: what is an extent of a detrimental influence 
of $U$ on the pairing correlations ? 

The abovementioned competition has been already studied 
in a normal phase of the BF model using the nonperturbative 
approach of the DMFT \cite{Romano-01}. In our paper we 
shall investigate the anisotropic superconducting phase. 
To make our study feasible we assume that the pairing 
interactions are relatively weak (a meaning of this assumption 
is explained in the next section) and we try to estimate 
the influence of the Coulomb repulsion varying its intensity 
from the weak to strong interaction limits.

\section{The model}
Hamiltonian of the system under consideration consists 
of the two parts $H=H^{BF} + H^{Coul}$. First of them 
refers to the standard BF model Hamiltonian \cite{Ranninger-95} 
\begin{eqnarray}
H^{BF} & = & \sum_{{\bf k},\sigma} \left( \varepsilon_{\bf k} 
- \mu \right) c_{{\bf k},\sigma}^{\dagger} c_{{\bf k},\sigma} 
+ \left( \Delta_{B} - 2\mu \right) \sum_{i} b_{i}^{\dagger} b_{i} 
\nonumber \\ & + &
\frac{1}{\sqrt{N}} \sum_{{\bf k},{\bf q}} \left( v_{{\bf k},{\bf q}} 
b_{\bf q}^{\dagger} c_{-{\bf k}+{\bf q}/2,\downarrow}
c_{{\bf k}+{\bf q}/2,\uparrow} + \mbox{h.c.} \right)
\label{BF}
\end{eqnarray}
and the second part denotes the on-site interaction between 
fermions $H^{Coul} = U \sum_{i} n_{i\downarrow} n_{i\uparrow}$.
We use here the standard notations for annihilation (creation) 
operators of fermion $c_{i,\sigma}$ ($c_{i,\sigma}^{\dagger}$) 
with spin $\sigma$ and for the hard-core boson $b_{i}$ 
($b_{i}^{\dagger}$) at site $i$ of the 2 dimensional square 
lattice. The indices ${\bf k}$ and ${\bf q}$ in (\ref{BF})
denote the coordinates of the momentum space. We assume the 
tight binding dispersion for fermions $\varepsilon_{\bf k}=-2t 
\left( \cos{k_{x}}+ \cos{k_{y}} \right)$, and set the bandwidth
$D=8t$ as a unit ($D \equiv 1$).

It is important to remark now that we let the boson-fermion 
exchange potential $v_{{\bf k},{\bf q}}$ to be anisotropic. 
As explained in the ref.\ \cite{Geshkenbeim-97} the low energy 
physics of this  model is prevailed by bosons of small momenta 
$|{\bf q}| \simeq 0$. Usually, the magnitudes of the superconducting 
gap in HTSC materials are of the order of several $meV$ (which is 
$\sim 10^{-3}$ of the bandwidth $D$). It is thus reasonable to assume 
that the pairing potential $v_{{\bf k},{\bf q}}$, which establishes 
the energy scale for $T_{c}$ and $\Delta_{sc}(T=0)$, is small 
as compared to $D$. In such a case the following mean field 
decoupling is justified
\begin{eqnarray}
& & \sum_{\bf q} v_{{\bf k},{\bf q}} 
b_{\bf q}^{\dagger} c_{-{\bf k}+{\bf q}/2,\downarrow}
c_{{\bf k}+{\bf q}/2,\uparrow} 
\simeq \nonumber \\ 
& & v_{{\bf k},{\bf q}={\bf 0}} 
\left( \left< b_{{\bf q}=\bf{0}} \right>^{*}
c_{ -{\bf k},\downarrow} c_{{\bf k},\uparrow} + 
b_{{\bf q}={\bf{0}}}^{\dagger}
\left< c_{ -{\bf k},\downarrow} c_{{\bf k},\uparrow} \right> 
\right) .
\label{decoupling}
\end{eqnarray} 
We further write down the anisotropic potential $v_{{\bf k},{\bf 0}}$
as a product \cite{Micnas-01}
\begin{equation}
v_{{\bf k},{\bf 0}} \equiv g \; \phi_{\bf k},
\label{BFpotential}
\end{equation}
where $g$ characterizes the interaction strength and $\phi_{\bf k}$
stands for the dimensionless factor which has to reflect the fourfold
symmetry  of CuO$_{2}$ planes of the HTSC cuprates. In general
the anisotropy factor $\phi_{\bf k}$ can be represented as
\begin{eqnarray}
\phi_{\bf k}= \alpha_{0} + \alpha_{s} \left( \cos{k_{x}} 
+ \cos{k_{y}} \right) + \alpha_{d} \left( \cos{k_{x}}
- \cos{k_{y}} \right)
\end{eqnarray}
and the coefficients $\alpha_{0,s,d}$ denote a relative 
contribution of the isotropic, the extended $s$-wave, and 
the $d$-wave part into the order parameter of the superconducting 
phase. They should be adjusted depending on a specific material.
If for example $\alpha_{0} \neq 0$ and $\alpha_{d} \neq 0$ we
would have the order parameter of a mixed $s+d$, or $s+$i$d$ 
symmetry. Since our main interest is focused on the competition 
between the Coulomb interaction and the superconductivity we
further consider for a clarity only the pure extended $s$-
or the $d$-wave symmetries when  $\phi_{\bf k}=\cos{k_{x}} \pm 
\cos{k_{y}}$. 

After the mean field decoupling for the boson-fermion
interaction (\ref{decoupling}) we are left with an effective
Hamiltonian composed of the separated fermion and boson 
contributions \cite{Ranninger-85} $H \simeq H^{F}+H^{B}$
\begin{eqnarray}
H^{F} & = & \sum_{{\bf k},\sigma} \xi_{\bf k}  
 c_{{\bf k},\sigma}^{\dagger} c_{{\bf k},\sigma} 
+ U \sum_{i} n_{i,\downarrow} n_{i,\uparrow}
\nonumber \\
& + & \sum_{\bf k} \left( g \; \rho \; \phi_{\bf k} 
c_{ {\bf k},\uparrow}^{\dagger} c_{-{\bf k},
\downarrow}^{\dagger} + {\mbox h.c.} \right) \;,
\label{H_F} \\
H^{B} & =  & \sum_{i}\left ( E_{0} b_{i}^{\dagger} b_{i} +
g \; x \; b_{i}^{\dagger} + g \; x^{*} b_{i} \right) \;.  
\end{eqnarray}
We introduced here the abbreviations for energies measured 
from the chemical potential $\xi_{\bf k}=\varepsilon_{\bf k}
-\mu$, $E_{0}=\Delta_{B}-2\mu$ and for the two order parameters
$x=\sum_{\bf k} \phi_{\bf k} \left< c_{ -{\bf k},\downarrow} 
c_{{\bf k},\uparrow} \right>$, $\rho=\left< b_{{\bf q}={\bf 0}} 
\right>/\sqrt{N}=\left< b_{i} \right>$.  

We can easily solve the hard-core boson part of the problem.
For a given site $i$ one finds the true eigenstates using
the unitary transformation
\begin{eqnarray}
|A>_{i} & = & \cos{ \left( \alpha \right)} |0>_{i} + 
\sin{\left(\alpha\right)} |1>_{i} \\
|B>_{i} & = & -\sin{\left(\alpha\right)} |0>_{i} + 
\cos{\left(\alpha\right)} |1>_{i}
\end{eqnarray}
such that $\tan{\left(2\alpha\right)}=(-2gx)/E_{0}$, where 
$|0>_{i}$, $|1>_{i}$ refer correspondingly to the empty and 
to the singly occupied (by the hard-core boson) site $i$. 
In a straightforward calculation we can determine the 
expectation values for the number operator $n^{B}=\sum_{i} 
\left< b_{i}^{\dagger} b_{i}\right>$ and for the order parameter 
$\rho$ \cite{Ranninger-95,Micnas-01}
\begin{eqnarray}
n^{B} & = & \frac{1}{2} - \frac{E_{0}}{4\gamma} 
\tanh{\left(\frac{\gamma}{k_{B}T}\right)}, \\
\rho & = & - \; \frac{gx}{2\gamma}\tanh{\left(
\frac{\gamma}{k_{B}T}\right)}
\end{eqnarray}
where $\gamma=\sqrt{(E_{0}/2)^{2}+|gx|^{2}}$ and $k_{B}$
is the Boltzmann constant.

\section{Weak interaction limit}
First, we consider the weak coupling limit when $U$ is 
fairly smaller than $D$. We are in a position to utilize 
then the Hartree Fock Gorkov linearization for
the on-site interaction $n_{i,\downarrow}n_{i,\uparrow} 
\simeq n_{i,\downarrow} \left< n_{i,\uparrow} \right> +
\left< n_{i,\downarrow} \right> n_{i,\uparrow} +  
\left< c_{i\uparrow}^{\dagger} c_{i,\downarrow}^{\dagger}
\right> c_{i,\downarrow}c_{i,\uparrow} +
c_{i\uparrow}^{\dagger} c_{i,\downarrow}^{\dagger}
\left< c_{i,\downarrow}c_{i,\uparrow} \right>$.
Hamiltonian of the fermion subsystem (\ref{H_F}) 
reduces then simply to the BCS structure
$H^{F}  \simeq  \sum_{{\bf k},\sigma} \tilde{\xi}_{\bf k}  
c_{{\bf k},\sigma}^{\dagger} c_{{\bf k},\sigma} 
+  \sum_{\bf k} \left( \Delta^{(eff)}_{\bf k} 
c_{{\bf k},\uparrow}^{\dagger} c_{ -{\bf k},
\downarrow}^{\dagger}  + {\mbox h.c.} \right) $
with $\tilde{\xi}_{\bf k}=\xi_{\bf k}+Un^{F}/2$ (we 
assume a paramagnetic state $\left< n_{i,\uparrow}\right> 
= \left< n_{i,\downarrow} \right> \equiv n^{F}/2$). A role 
of the effective gap parameter is played here by
\begin{eqnarray}
\Delta^{(eff)}_{\bf k} = \Delta_{0} + g \; \rho \; \phi_{\bf k},
\label{eff_order}
\end{eqnarray}
where the isotropic part is given by $\Delta_{0}= 
U \left< c_{i,\downarrow}c_{i,\uparrow} \right>$.
Standard methods of the solid state theory give the following
equations for expectation values
\begin{eqnarray}
n^{F} & = & 1 - \sum_{\bf k} \frac{\tilde{\xi}_{\bf k}}
{E_{\bf k}} \; \tanh{\left(\frac{E_{\bf k}}{2k_{B}T}\right)} \;, 
\label{n_F} \\
\left< c_{-{\bf k},\downarrow} c_{{\bf k},\uparrow} \right>
& = & \frac{-\Delta_{\bf k}^{(eff)}}{2E_{\bf k}}
\; \tanh{\left( \frac{E_{\bf k}}{2k_{B}T} \right) } \;,
\label{ck_ck}
\end{eqnarray}
with a typical gaped spectrum $E_{\bf k}=\sqrt{\tilde
{\xi}_{\bf k}^{2} + | \Delta_{\bf k}^{(eff)}|^{2}}$ 
in the superconducting phase.

It is worth mentioning that in a case of the $d$-pairing
(i.e.\ for $\phi_{\bf k} =\cos{k_{x}} - \cos{k_{y}}
\equiv \eta_{\bf k}$) the isotropic component $\Delta_{0}$
of the gap parameter (\ref{eff_order}) does identically
vanish. To prove this let us substitute (\ref{ck_ck}) 
into the definition of $\Delta_{0}= U \sum_{\bf k} \left< 
c_{-{\bf k},\downarrow} c_{{\bf k},\uparrow} \right>$ to
obtain
\begin{eqnarray}
\Delta_{0}  =  - U \sum_{\bf k} \frac{\Delta_{0}
+g \; \rho \; \eta_{\bf k}}{2E_{\bf k}} \; \tanh{\left( 
\frac{E_{\bf k}}{2k_{B}T}\right) } \;.
\label{Delta_0d}
\end{eqnarray} 
Since integration over the Brillouin zone of the part 
containing $\eta_{\bf k}$ gives zero so, for $U>0$,  
equation (\ref{Delta_0d}) has the only possible solution 
$\Delta_{0}=0$. It is not surprising because the repulsive 
interactions by themselves are not able to induce the 
on-site fermion pairs. 

If the boson fermion potential (\ref{BFpotential}) is
isotropic or takes a form of the $s$-wave ($\phi_{\bf k} 
=\cos{k_{x}} + \cos{k_{y}}$) then in general $\Delta_{0}
\neq 0$. From a consideration similar to the one discussed 
above (equation (\ref{Delta_0d}) is valid except that
$\eta_{\bf k}$ should be replaced by $\phi_{\bf k}$) we 
can determine a relative ratio $\Delta_{0}/g\rho$. 
The extended $s$-wave gap parameter is now given by
\begin{eqnarray}
& &\Delta_{\bf k}^{(eff)} =  \nonumber \\
& & g \; \rho \left(  \phi_{\bf k}   -
\frac{\sum_{\bf k} \left( U\phi_{\bf k}/2E_{\bf k} \right)
\tanh{ \left( E_{\bf k}/2k_{B}T \right)} }{1+
 \sum_{\bf k} \left( U/2E_{\bf k} \right)
\tanh{ \left( E_{\bf k}/2k_{B}T \right)} } \right)  \;.
\label{extended-s}
\end{eqnarray}
For the isotropic boson fermion potential 
($\phi_{\bf k}=1$) the equation (\ref{extended-s}) 
simplifies further to give a ${\bf k}$-independent 
gap $\Delta^{(eff)} =  g \rho \left[ 1+\sum_{\bf k} 
\left( U/2E_{\bf k} \right)\tanh{ \left( E_{\bf k}
/2k_{B}T \right)} \right]^{-1} < g\rho$. This expression
explicitly shows a detrimental role of the on-site
repulsion on the isotropic superconducting phase. 
Such a problem has been previously  addressed
\cite{Kostyrko-96,Kostyrko} neglecting the hard core 
nature of bosons and using the RPA treatment for 
the Coulomb repulsion.

\section{Strong interaction limit}
In a case of the  strong interactions  ($U > D$) 
we make use of the slave boson technique proposed 
by Kotliar and Ruckenstein \cite{Kotliar-86}. For 
simplicity we shall consider here only the extreme 
limit $U \rightarrow \infty$.  

We represent the fermion operators as $c_{i,\sigma} = 
a_{i}^{\dagger} f_{i,\sigma}$ and $c_{i,\sigma}^{\dagger} 
= f_{i,\sigma}^{\dagger}a_{i}$, where the auxiliary boson
operator $a_{i}$ ($a_{i}^{\dagger}$) refers to the annihilation
(creation) of the empty state at site $i$ and fermion operator 
$f_{i,\sigma}$ ($f_{i,\sigma}^{\dagger}$) corresponds
to annihilation (creation) of the singly occupied site $i$ with
spin $\sigma$. No double occupancy is allowed and this
can be formally expressed via the local constraint 
$a_{i}^{\dagger}a_{i} + \sum_{\sigma} f_{i,\sigma}^{\dagger}
f_{i,\sigma}=1$. 

Using the real space (Wannier states) representation we can 
rewrite the Hamiltonian (\ref{H_F}) in terms of the new 
operators as
\begin{eqnarray}
H^{F} & = & \sum_{i,j,\sigma} t_{i,j} f_{i,\sigma}^{\dagger}
a_{i} a_{j}^{\dagger} f_{j,\sigma} - \mu \sum_{i,\sigma}
f_{i,\sigma}^{\dagger}f_{i,\sigma} 
\nonumber \\ & + &
\left( \rho \sum_{i,j} V_{i,j} f_{i,\uparrow}^{\dagger} 
a_{i} a_{j} f_{j,\downarrow}^{\dagger} + \mbox{h.c.} \right) 
\nonumber \\ & + &
\sum_{i} \lambda_{i} \left(  a_{i}^{\dagger}a_{i} + \sum_{\sigma}
f_{i,\sigma}^{\dagger}f_{i,\sigma} -1 \right) \;.
\label{HF_sb}
\end{eqnarray}
We used here the identity $a_{i}a_{i}^{\dagger}=1$ 
\cite{Kotliar-86} and the last term takes account of the
local constraint ($\lambda_{i}$ stands for the Lagrange
multiplier). $V_{i,j}$ is the exchange potential whose 
Fourier transform is given by (\ref{BFpotential}) and, 
as usually, $t_{i,j}$ denotes the hopping integral.  

Next, we approximate (\ref{HF_sb}) by: (i) replacing 
the slave boson operators by their expectation values 
which are assumed to be site independent $a_{i} \simeq  
\left< a_{i} \right> \simeq r$, and (ii) replacing the
local multipliers by the global one $\lambda_{i} \simeq
\lambda$. In this (mean field) approximation for the slave
bosons one obtains
\begin{eqnarray}
H^{F} & \simeq & \sum_{{\bf k},\sigma} \left( 
r^{2} \varepsilon_{\bf k} - \mu + \lambda \right)  
f_{{\bf k},\sigma}^{\dagger} f_{{\bf k},\sigma} 
\nonumber \\
& + & \sum_{\bf k} \left( r^{2}g \; \rho \; \phi_{\bf k} 
f_{ {\bf k},\uparrow}^{\dagger} f_{-{\bf k},
\downarrow}^{\dagger} + {\mbox h.c.} \right) \;.
\label{sb_HF_mf}
\end{eqnarray} 
The global parameters $\lambda$, $r$ are determined from 
a minimization of the total energy $\left< H \right>$. 
This criterion leads to
\begin{eqnarray}
r^{2} & = & 1-n^{F} ,\\
\lambda & = & - \sum_{{\bf k},\sigma} \varepsilon_{\bf k} 
\left< f_{{\bf k},\sigma}^{\dagger} f_{{\bf k},\sigma} \right> 
\nonumber \\ & & -2 {\rm Re} \left\{ g \rho^{*}  
\sum_{\bf k} \phi_{\bf k} \left< f_{-{\bf k},\downarrow} 
f_{{\bf k},\uparrow} \right> \right\} \;.
\end{eqnarray} 

As can be noticed from the equation (\ref{sb_HF_mf}) Hamiltonian
of the fermion subsystem $H^{F}$ is again reduced to the BCS 
structure. We thus have the same solution for the expectation 
values as given in (\ref{n_F},\ref{ck_ck}) with a difference 
that now
\begin{eqnarray}
\tilde{\xi}_{\bf k} & = &  r^{2}\varepsilon_{\bf k} - \mu
+ \lambda  \;, \\
\Delta_{\bf k} & = & r^{2} \; g\; \rho \; \phi_{\bf k} \;.
\end{eqnarray}
Both, the effective fermion bandwidth $D^{(eff)}=r^{2}D$
and the effective pairing  potential $V_{\bf k}^{(eff)}=r^{2}
g\phi_{\bf k}$, reduce down to zero when fermion occupation
approaches the half-filling. Under such circumstances the system 
is driven into the Mott insulating state.

\section{The crossover}
Finally we consider here a regime of the intermediate $U$ for 
which we adopt the procedure used earlier by us \cite{TD_KIW-99}
in a context of the extended Hubbard model. We introduce
the Nambu representation $\Psi_{\bf k}^{\dagger}=
(c_{{\bf k},\uparrow}^{\dagger} c_{-{\bf k},\downarrow})$, 
$\Psi_{\bf k}=(\Psi_{\bf k}^{\dagger})^{\dagger}$
and, as a first step, determine the unperturbed Green's
function $G^{0}({\bf k},\omega)=\left<\left<\Psi_{\bf k};
\Psi_{\bf k}^{\dagger}\right>\right>_{\omega}$ neglecting 
the interaction $U$ in the fermion Hamiltonian (\ref{H_F})  
\begin{eqnarray} 
\left[ \begin{array}{cc} 
i \omega_{n} - \xi_{\bf k}  
~~~~~- g \rho^{*} \phi_{\bf k} \\
- g \rho \phi_{\bf k} 
~~~~~i \omega_{n} - \xi_{\bf k} 
\end{array} \right] \; 
G_{0}({\bf k},i\omega_{n})  = {\bf 1} \;.
\end{eqnarray}  
Next, we compute the dressed Green's function
using the matrix Dyson equation
\begin{eqnarray}
G^{-1}({\bf k},i\omega_{n})=
G_{0}^{-1}({\bf k},i\omega_{n}) 
-\Sigma({\bf k},i\omega_{n}) \;.
\end{eqnarray}
In order to proceed we simplify the selfenergy matrix
by the following Ansatz \cite{TD_KIW-99}
\begin{eqnarray}
\Sigma({\bf k},i\omega_{n}) \simeq
\left[ \begin{array}{rl} 
\Sigma_{N}({\bf k},i\omega_{n}) & U \left< 
c_{i\uparrow}^{\dagger} c_{i\downarrow}^{\dagger}
\right> \\
U \left< c_{i\downarrow} c_{i\uparrow} \right>
& -\Sigma_{N}({\bf k},-i\omega_{n}) 
\end{array} \right] \;.
\label{Sigma}
\end{eqnarray}
Without specifying the diagonal elements of (\ref{Sigma}) 
we denote $\Sigma_{11}({\bf k},i\omega_{n})$ by $\Sigma_{N}
({\bf k},i\omega_{n})$ and make use of the identity 
$\Sigma_{22}({\bf k},i\omega_{n})=-\Sigma_{11}({\bf k},
-i\omega_{n})$. The off-diagonal elements are approximated
by us by a contribution corresponding to the result 
deduced from the mean field type theory discussed in 
the section III. Channel of the superconducting correlations 
is treated by us in (\ref{Sigma}) approximately.    

If one knew $\Sigma_{N}({\bf k},i\omega_{n})$
then the needed expectation values can be found 
according to the standard field theoretical relation 
$\left< A B \right> = \beta^{-1} \sum_{n=\infty}^{\infty} \left< 
\left< B; A \right> \right> _{i\omega_{n}}$, where
$\beta=1/k_{B}T$. In particular, we obtain \cite{TD_KIW-99}
\begin{eqnarray}
\left< c_{{\bf k}\uparrow}^{\dagger} c_{{\bf k}\uparrow}
\right> & = & \beta^{-1} \sum_{n} \frac{i\omega_{n}+\xi_{\bf k} 
+ \Sigma_{N}(-i\omega_{n})}{|i\omega_{n}-\xi_{\bf k} - 
\Sigma_{N}(i\omega_{n})|^{2}+|\Delta^{(eff)}_{\bf k} |^{2}}
\nonumber \\ & & \\ 
\left< c_{-{\bf k}\downarrow} c_{{\bf k}\uparrow}\right> 
& = & \beta^{-1} \sum_{n} \frac{\Delta^{(eff)}_{\bf k}}
{|i\omega_{n}-\xi_{\bf k} - \Sigma_{N}(i\omega_{n})|^{2}
+|\Delta^{(eff)}_{\bf k} |^{2}} ,
\nonumber \\ & &
\end{eqnarray} 
where again $\Delta_{\bf k}^{(eff)}$ is given by
(\ref{eff_order}).

It is worth noticing that $\Sigma_{N}({\bf k},i\omega_{n})$ 
has a meaning of the normal phase selfenergy for the standard 
Hubbard model. Of course there is no exact solution for 
$\Sigma_{N}$ available so far except maybe from the numerical
exact diagonalization or the Quantum Monte Carlo studies. 
However, depending on a magnitude of the on-site interaction, 
one can use various approximate estimations. Let us point out 
the few possibilities.

Starting from the weak interaction limit the simplest
substitution for $\Sigma_{N}$ is the mean field value 
$U n^{F}/2$ as discussed in section III. With an increase of 
$U$ one can proceed by including some higher order corrections, 
like for example of the second order in $U$ \cite{SOPT}.
Going toward the Mott transition regime $U=U_{cr} \sim D$ 
(for the half-filled fermion system) one could work for example 
with the so called alloy analogy approximation (AAA) 
\cite{TD_KIW-99}. In a more subtle way one could estimate 
the momentum independent selfenergy $\Sigma_{N}(\omega)$ 
by adopting  the dynamical mean field theory (which becomes 
exact in the limit of infinite spatial dimensions). The strong 
interaction case can be described in a satisfactory way either 
with AAA, DMFT or the simple Hubbard I approximation. Here we 
apply the AAA procedure to study a regime near the Mott transition.

\section{Numerical results}
As far as the Coulomb repulsion is concerned we 
expect that its effectiveness should strongly depend on 
concentration of fermions. In a regime of small fermion
concentration (dilute limit) this interaction should not
be very efficient and, in particular, it would not 
affect the superconducting type correlations induced
by the boson fermion exchange. On the other hand,
we expect that the strongest effects of the on-site
Coulomb repulsion might appear for a system
with a nearly half filled fermion system $n_{F} =1$.

In absence of the Coulomb interactions, the superconducting
phase (isotropic or anisotropic one) of the BF model is 
formed when the fermion concentration is properly adjusted.
The Fermi energy $\varepsilon_{F}$
must be close enough to the boson level because only then
the charge exchange between the hard-core bosons and fermion
pairs can induce the long range coherence \cite{Ranninger-90}.
So, the needed concentration of fermions is roughly
given by $n_{F}=
2 \int_{-D/2}^{\varepsilon_{F}=\Delta_{B}/2} \rho_{0}
(\varepsilon)d\varepsilon$, where $\rho_{0}(\varepsilon)$ 
denotes the density of states of the free (noninteracting)
fermion system.

\begin{figure}
\centerline{\epsfxsize=8cm \epsfbox{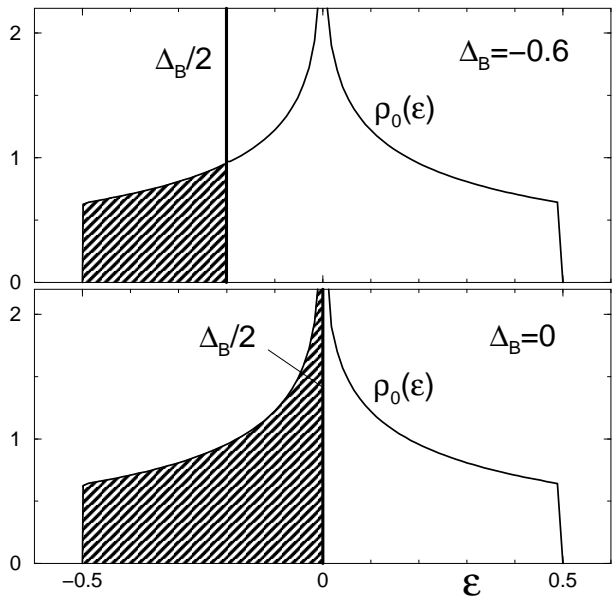}}
\vspace{3mm}
\caption{A schematic illustration of the two distinct
critical fermion concentrations: (top) $n_{F} \simeq 0.4$
when boson energy $\Delta_{B}/2=-0.3$, and (bottom)
$n_{F}=1$ when $\Delta_{B}/2=0$. The shaded areas show
the occupied fermion states at zero temperature.
}
\label{demo}
\end{figure}

In this section we present results obtained numerically
for the BF model in the two distinct cases when
the critical fermion concentration is:
a) small, which takes place when $\Delta_{B}/2$
lies fair aside the center of fermion band,
b) close to half-filling, when $\Delta_{B}/2$ is
located in the center of fermion band. We thus
choose the two following values $\Delta_{B}/2=-0.3$
and $0$, see figure (\ref{demo}).

\begin{figure}
\centerline{\epsfxsize=8cm \epsfbox{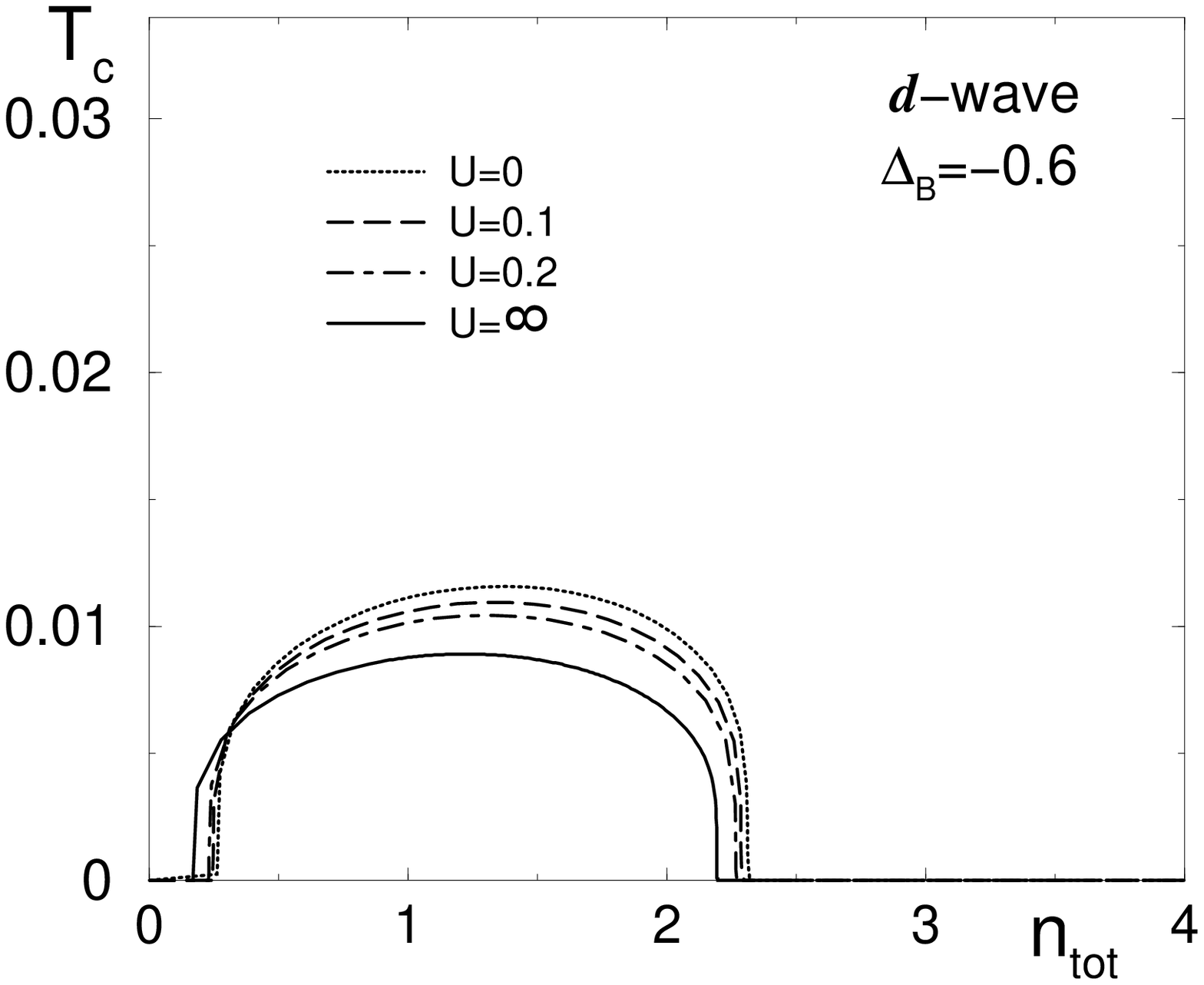}}
\centerline{\epsfxsize=8cm \epsfbox{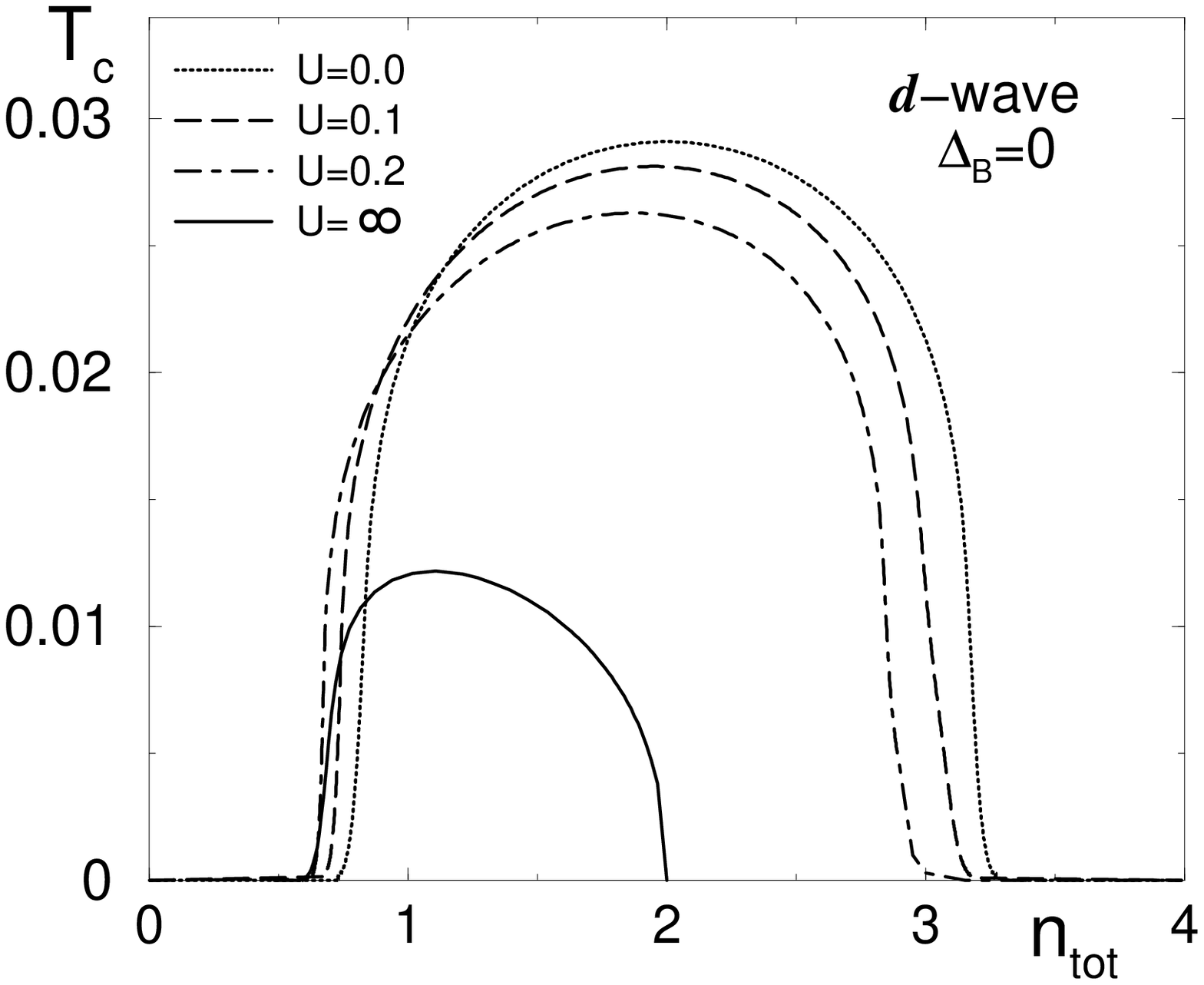}}
\vspace{3mm}
\caption{
Transition temperature $T_{c}$ into the $d$-wave
superconducting phase as a function of the total concentration
$n_{tot}$ for the two representative boson level values $\Delta_{B}=-0.6$
(top) and $\Delta_{B}=0$ (bottom). Curves corresponding to $U=0.1$
and $0.2$ where obtained from the mean field approximation,
while $U=\infty$ from the salve boson study. }
\label{fig1}
\end{figure}

We take the boson fermion potential $g=0.1$ in all
the results discussed below. Figure (\ref{fig1}) shows
the critical temperature $T_{c}$ of the $d$- wave
superconducting phase for several values of $U$.
This type of superconductivity is enhanced
near the half-filled fermion system similarly as in
the extended Hubbard model \cite{Ranninger-90}.
In agreement to our expectations, the Coulomb repulsion
only weakly reduces $T_{c}$ in a dilute fermion system.
However, for $\Delta_{B}=0$ we notice a considerable
reduction of $T_{c}$ or even a disappearance of
superconductivity for $n_F=1$ (total concentration
is then $n_{tot}=n_{F}+2n_{B}=2$) when $U$ exceeds
the critical Mott transition value $U_{cr}$. For such a
strong potential $U$ the $d$-wave superconducting phase
is restricted to a narrower regime of the total concentration,
such that there are no doubly occupied fermion states
on a given lattice site (remember we are considering
the intersite Cooper pairs).

\begin{figure}
\centerline{\epsfxsize=8cm \epsfbox{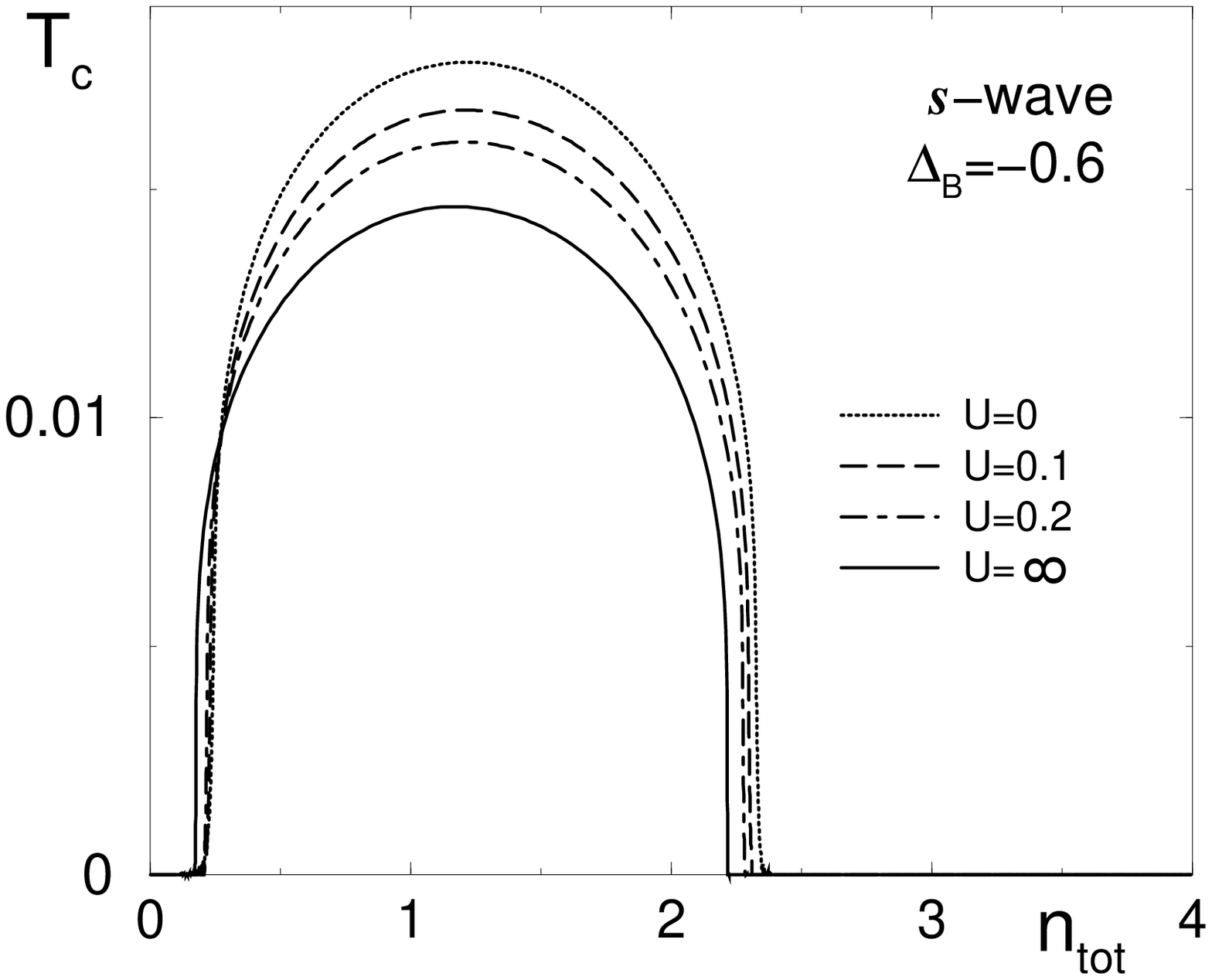}}
\centerline{\epsfxsize=8cm \epsfbox{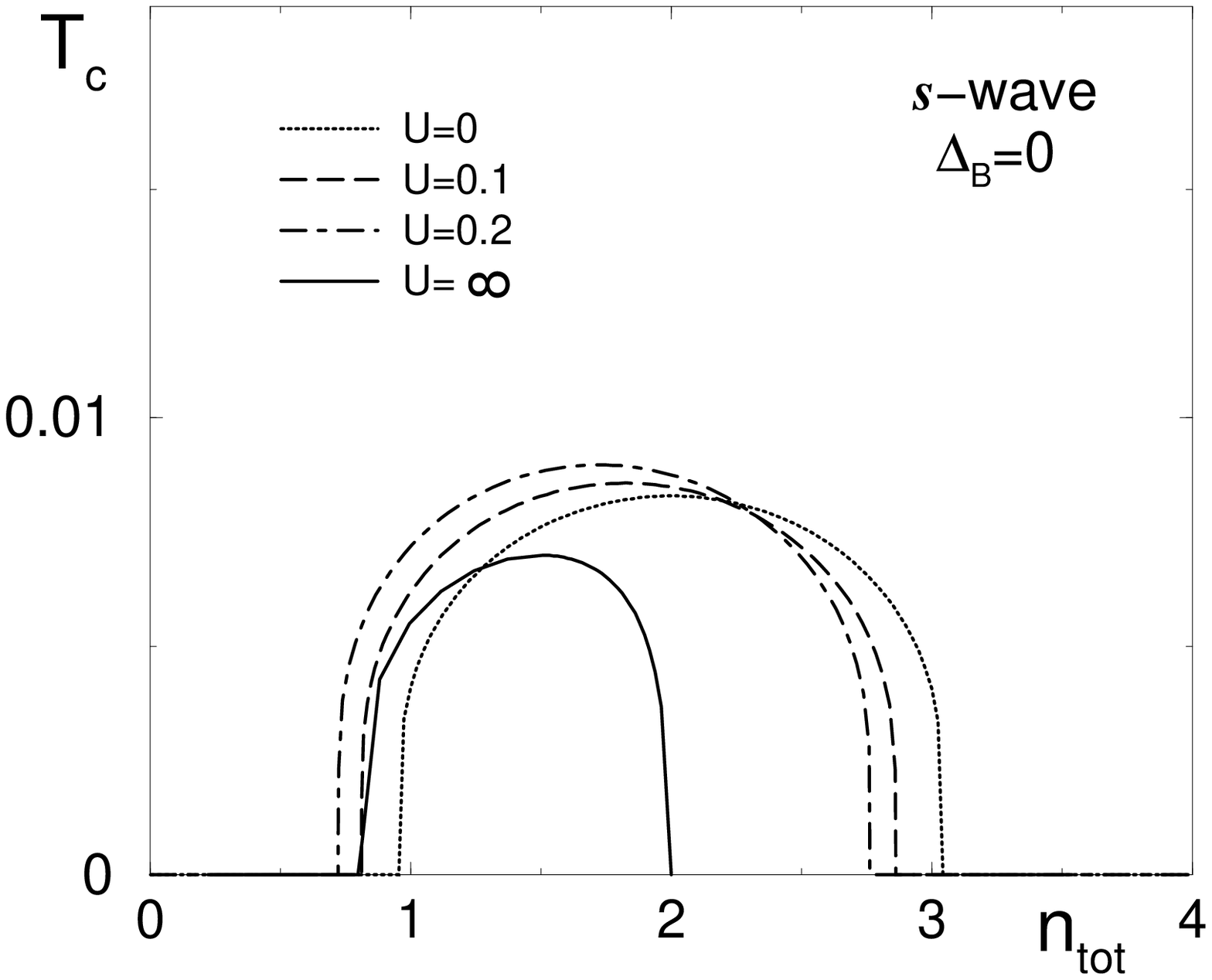}}
\vspace{3mm}
\caption{
Transition temperature $T_{c}$ into the extended $s$-wave 
superconducting phase for several values of the on-site 
repulsion, as indicated. The top figure corresponds to 
$\Delta_{B}=-0.6$ and the bottom one to $\Delta_{B}=0$.
}
\label{fig2}
\end{figure}

Figure (\ref{fig2}) illustrates the effect of $U$ on 
the extended $s$-wave phase. In a dilute regime
we notice almost identical values of $T_{c}$
for both the $d$- and extended $s$-wave superconducting
phases. Also the influence of $U$ is there very similar. 
A remarkable difference appears for $\Delta_{B}=0$ when
the fermion system is near the half-filling $n_{F}=1$.
Critical temperature $T_{c}$ of the $s$-wave phase is
then 3 times smaller as compared to the $d$-wave phase. 
The system is thus less susceptible for the $s$ type pairing 
near $n_{F}=1$ (for $\Delta_{B}=$ it corresponds to $n_{tot}=2$). 

One notices also some "peculiar" behavior of $T_{c}(n_{tot})$
for the extended $s$-wave phase cased by an increasing 
strength of $U$. With a small increase of $U$ the whole
diagram is somewhat shifted and simultaneously the optimal
value of $T_{c}$ slightly increases. This overall shift
is caused by the Hartree term $Un_{F}/2$ (see the section
III) which effectively pulls up the fermion band with 
respect to the boson energy level. By comparing the curves 
corresponding to $U=0$ in the upper and bottom panels of
figure (\ref{fig2}) we realize that such shift is responsible 
for enhancing the $s$-wave type superconductivity, but only 
when $U$ is safely smaller than $D$. A further increase of 
the Coulomb interaction $U$ proves to be detrimental on
superconductivity (independently of a symmetry of the order 
parameter) because the fermion subsystem is driven into
the Mott insulating state.

\begin{figure}
\centerline{\epsfxsize=8cm \epsfbox{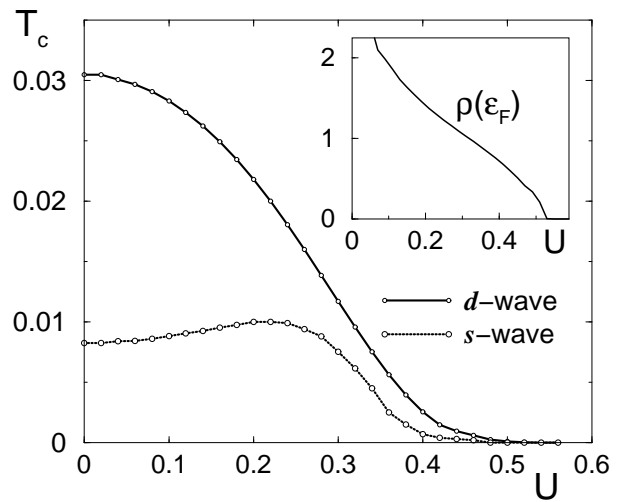}}
\vspace{3mm}
\caption{
Transition temperature $T_{c}$ of the $d$- and the extended 
$s$-wave phases for the half-filled fermion system $n_{F}=1$. 
Both superconducting phases disappear when the Mott transition 
is approached. The inset shows the density of states at the Fermi 
energy for a normal phase obtained by the Alloy Analogy
Approximation.  
}
\label{fig3}
\end{figure}

To analyze in more detail a pronounced effect of
the Coulomb interaction on superconductivity
for the half-filled fermion system $n_{F}=1$ 
($n_{tot}=2$) we show in figure (\ref{fig3}) 
the dependence of $T_{c}$ on $U$. The results have 
been obtained by means of the Alloy Analogy Approximation 
mentioned in section III and discussed earlier
by the same author in Ref.\ \cite{TD_KIW-99}. 
For the $dim=2$ tight binding dispersion we determine
the Mott transition at $U_{cr} \simeq 0.54$ in units
of the initial fermion band. This value is probably
underestimated. The most credible determination
based on the dynamical mean field theory usually
yields $U_{cr}$ larger than 1 \cite{DMFT-review}.
Nevertheless, a qualitative behavior presented in
figure (\ref{fig3}) remains valid. As we see by 
comparing to the figures (\ref{fig1},\ref{fig2})
the AAA treatment properly interpolates
between $U=0$ and $U=\infty$ limits.

\section{Conclusions}
Summarizing, we have investigated the anisotropic 
superconductivity within the boson fermion model 
in a presence of the Coulomb repulsion between 
fermions. 
\\
(a) In a dilute regime of the fermion concentration 
the effect of the Coulomb repulsion proves to be rather 
weak. For both the $d$- and extended $s$-wave phase we 
observe up to 25 $\%$ reduction of the optimal $T_{c}$ 
value when $U \rightarrow \infty$. Both anisotropic phases 
survive, even in the limit of infinitely strong Coulomb 
repulsions. 
\\
(b) In the nearly half-filled fermion system
we observe an enhancement of the $d$-wave superconducting 
phase and a simultaneous suppression of the $s$-wave 
phase until the interaction $U$ is small. 
\\
(c) Around the Mott transition $U_{cr} \simeq 0.54$
both phases are reduced to the concentration regime 
$n_{F}<1$, $2n_{B}<1$. Still, the superconductivity 
is able to survive at sufficiently large
hole concentration $h=1-n_{F}>0$. Such a case
is relevant for a description of the HTSC materials
and boson fermion model seems to capable to 
reproduce qualitatively the phase diagrams
known for these materials.  

Among the problems which are not addressed in this
paper there is a very intriguing question: what happens
to pseudogap of the normal phase (discussed earlier
in the Refs \cite{Domanski-01,perturbative,Ren-98,DMFT}
and in \cite{Micnas-01} in a presence of Coulomb
interactions ? Consideration of this subject is in a
progress and the results shall be discussed elsewhere.

\acknowledgements{
Author kindly acknowledges helpful discussions with J.~Ranninger
and K.I.~Wysoki\'nski. Partial support has been provided by 
the Polish Committee of Scientific Research under project 
No.\ 2P03B 106 18.}


\begin{references}
\bibitem{Ranninger-85}
    J.~Ranninger and S.~Robaszkiewicz, Physica {\bf B 135}, 468 (1985).
\bibitem{Eliashberg-87}
    G.M.~Eliashberg, Pis'ma Zh.~Eksp.~Teor.~Fiz.\ {\bf 46}, 94 (1987).    
\bibitem{Friedberg-89}
    R.~Friedberg and T.D.~Lee, Phys.~Rev.\  {\bf B 40}, 423 (1989);
    R.~Friedberg, T.D.~Lee and H.C.~Ren, Phys.~Lett.~A {\bf 152},
    417 (1991).
\bibitem{Ioffe-89}
    L.~Ioffe, A.I.~Larkin, Y.N.~Ovchinnikov and L.~Yu, 
    Int.~Journ.~Modern~Phys.\ {\bf B 3}, 2065 (1989).    
\bibitem{Ranninger-95}    
    J.~Ranninger and J.M.~Robin, Physica {\bf C 253}, 279 (1995). 
\bibitem{Holland-01}
    M.~Holland, S.J.J.M.F.~Kokkelmans, M.L.~Chiofalo
    and R.~Walser, Phys.~Rev.~Lett.\ {\bf 87}, 120406 (2001).    
\bibitem{Kostyrko-96}
    T.~Kostyrko and J.~Ranninger, Phys.~Rev.\ {\bf B 54}, 13105 (1996).
\bibitem{Domanski-01}
    T.~Doma\'nski and J.~Ranninger, Phys.~Rev.\ {\bf B 63}, 134505 (2001).
\bibitem{Timusk-99}
    T.~Timusk and B.~Statt, Rep.~Prog.~Phys.\ {\bf 62}, 61 (1999).
\bibitem{perturbative}
    J.~Ranninger, J.M.~Robin, M.~Eschrig, Phys.~Rev.~Lett.\ {\bf 74}, 
    4027 (1995);
    J.~Ranninger and J.M.~Robin, Solid State Commun.\
    {\bf 98}, 559 (1996);
    J.~Ranninger and J.M.~Robin, Phys.~Rev.\ {\bf B 53}, R11961 (1996);
    P.~Devillard and J.~Ranninger, Phys.~Rev.~Lett.\ {\bf 84}, 
    5200 (2000).
\bibitem{Ren-98}
    H.C.~Ren, Physica {\bf C 303}, 115 (1998).    
\bibitem{Domanski-98}
    T.~Doma\'nski, J.~Ranninger and J.M.~Robin, Solid State Commun.\
    {\bf 105}, 473 (1998).
\bibitem{DMFT}
    J.M.~Robin, A.~Romano, J.~Ranninger, Phys.~Rev.~Lett.\
    {\bf 81}, 2755 (1998); 
    A.~Romano and J.~Ranninger, Phys.~Rev.\ {\bf B 62}, 4066 (2000).
\bibitem{ARPES}
    H.~Ding {\em et al}, Nature {\bf 382}, 51 (1996);
    A.G.~Loeser {\em et al}, Science {\bf 273}, 325 (1996).
\bibitem{Kouznetsov-97}
    K.A.~ Kouznetsov {\em et al}, Phys.~Rev.~Lett.\ 
    {\bf 79}, 3050 (1997);
    A.G.~Sun {\em et al}, Phys.~Rev.~Lett.\ {\bf 72}, 2267 (1995).
\bibitem{Ma-95}
    J.~Ma {\em et al}, Science {\bf 267}, 862 (1995);
    H.~Ding, J.C.~Campuzano and G.~Jennings, Phys. Rev. Lett.
    {\bf 74}, 2784 (1995).
\bibitem{Musaelian-96}
    J.~Betouras and R.~Joynt, Europhys.~Lett.\ {\bf 31}, 119 (1995);
    K.A.~Musaelian, J.~Betouras, A.V.~Chubukov and R.~Joynt,
    Phys.~Rev.\ {\bf B 53}, 3598 (1996).
\bibitem{Geshkenbeim-97}
    V.B.~Geshkenbein, L.B.~Ioffe and A.I. Larkin, Phys.~Rev.\
    {\bf B 55}, 3173 (1997).
\bibitem{Micnas-01}
    R.~Micnas, S.~Robaszkiewicz and B.~Tobijaszewska,
    Physica {\bf B 312}-{\bf 313}, 49 (2002);
    R.~Micnas and B.~Tobijaszewska, Acta Phys. Pol.\
    {\bf B 32}, 3233 (2001).
\bibitem{Romano-01}
    A.~Romano, Phys.~Rev.\ {\bf B 64}, 125101 (2001).
\bibitem{Kostyrko}
    T.~Kostyrko, Acta Phys.~Pol.\ {\bf A 91}, 399 (1997).
\bibitem{Kotliar-86}
    G.~Kotliar and A.E.~Ruckenstein, Phys.~Rev.~Let.\ {\bf 57},
    1362 (1986);
    D.H.~Newns and N.~Read, Adv.~Phys.\ {\bf 36}, 799 (1987).
\bibitem{TD_KIW-99}
    T.~Doma\'nski and K.I.~Wysoki\'nski, Phys.~Rev.\
    {\bf B 59}, 173 (1999).
\bibitem{SOPT}
    H.~Schweitzer and G.~Czycholl, Z.~Phys. {\bf B 83}, 93 (1991).
\bibitem{Ranninger-90}
    R.~Micnas, J.~Ranninger and S.~Robasz\-kie\-wicz,
    Rev. Mod. Phys. {\bf 62}, 113 (1990)
\bibitem{DMFT-review}
    A.~Georges, G.~Kotliar, W.~Krauth and M.~Rozenberg,
    Rev.~Mod.~Phys.\ {\bf 68}, 13 (1996).
\end{references}
\end{document}